\documentclass{PoS}

\usepackage[colorinlistoftodos]{todonotes}

\title{Studying the Temporal Variation of the Cosmic-Ray Sun Shadow Using IceCube Data}

\ShortTitle{Cosmic-ray Sun Shadow with IceCube}

\author{
The IceCube Collaboration\footnote{For collaboration list, see PoS(ICRC2019) 1177.}\\
{\itshape \href{http://icecube.wisc.edu/collaboration/authors/icrc19_icecube}{http://icecube.wisc.edu/collaboration/authors/icrc19\_icecube}}\\
E-mail: \email{frederik.tenholt@rub.de, julia.tjus@rub.de, paolo.desiati@icecube.wisc.edu}
}

\abstract{The shadowing effect of the Moon and Sun in
TeV cosmic rays has been measured with high statistical significance by
several experiments. Unlike particles from directions close to the Moon,
however, charged particles passing by the neighborhood of the Sun are affected not only by the
geomagnetic but also by the solar near- and interplanetary-magnetic field. Since the latter undergoes
a well-known 11-year cycle -- during which it can become highly disordered --
the cosmic-ray shadow cast by the Sun as observed on Earth is expected to change over time. We present an update of the analysis of the cosmic-ray Moon and
Sun shadows using data taken with the IceCube Neutrino Observatory. With
a median energy after quality cuts of approximately $50$--$60$~TeV, depending on the cosmic-ray flux model used,
primary cosmic rays inducing events which pass IceCube's Sun
shadow filter have a comparatively high energy. While the results for the
Moon shadow confirm the stability of the IceCube observatory, the results for the Sun shadow exhibit a clear variation correlating with solar activity and theoretical models of the solar magnetic field.

\vspace{4mm}
{\bfseries Corresponding authors:}
Frederik Tenholt$^{1,2}$, Julia Becker Tjus$^{1,2}$, \speaker{Paolo Desiati}$^{3}$\\
{$^{1}$ \itshape Institute for Theoretical Physics, Plasma-Astroparticle Physics, Ruhr-University Bochum, 44780 Bochum, Germany}\\
{$^{2}$ \itshape RAPP Center, Ruhr-University Bochum, 44780 Bochum, Germany}\\
{$^{3}$ \itshape WIPAC, Madison, Wisconsin, USA}
}

\FullConference{36th International Cosmic Ray Conference -ICRC2019-\\
		July 24th - August 1st, 2019\\
		Madison, WI, U.S.A.}

\begin{document}

\section{Introduction}\label{sec:info}
\subsection{Measurements of the cosmic-ray shadow of the Sun}
The Moon and the Sun block cosmic rays on their way to Earth. This shadowing effect has been measured with high statistical significance by several experiments like Tibet \cite{Amenomori-etal-2013,Amenomori-etal-2018prl,Amenomori-etal-2018apj} and Argo \cite{argo2017}.
The Moon serves as an absolute pointing calibration,
 meaning that it shows that the direction of the shadow is indeed where the Moon is expected to be. The Sun shadow, on the other hand, serves as an indirect measurement of the solar magnetic field, which has not yet been measured directly, close (few solar radii distance) to the Sun. Previously, the Tibet Collaboration proved that the cosmic-ray shadows varies with the 11-year cycle of the solar magnetic field \cite{Amenomori-etal-2013} and that the observed shift of the cosmic-ray shadow requires an interplanetary magnetic field that is at least a factor of $1.5$ larger than measured with indirect B-field detections \cite{Amenomori-etal-2018prl}. The influence of Coronal Mass Ejections (CMEs) on the shadow could be proven in \cite{Amenomori-etal-2018apj}. All Tibet measurements are sensitive to energies at $\sim10$~TeV. ARGO-YBJ investigates the shadow in a rigidity range of $\sim 0.2 - 200$~TV, proving the variation of the shadow up to at least $100$~TV. The study presented here investigates cosmic rays with median energies of $\sim 48-85$~TeV for the moon and $\sim 50-60$~TeV for the Sun. The exact numbers depend on the primary model in both cases and on the exact setting of the season for the moon.
%\textbf{frederik, number ok?} \textcolor{red}{Frederik: For the Moon, the median energy is between 48 and 85 TeV, depending on a) the season and b) the primary model. For the Sun, it is between 50 and 60 TeV, depending on the primary model (no change over the seasons). So we can either give that range (50-60TeV), or say approximately 55 TeV. I also slightly changed the abstract to be consistent with this}. 

\subsection{IceCube as a cosmic ray detector}
IceCube is a cubic-kilometer neutrino detector installed in the Antarctic ice at the geographic South Pole \cite{Aartsen:2016nxy} between depths of 1450 m and 2450 m, completed in 2010. 
It consists of 86 strings with a total number of 5160 digital optical modules (DOMs) that detect light emitted from charged particles traveling through the ice.  
Reconstruction of the direction, energy and flavor of the neutrinos therefore relies on the optical detection of Cherenkov radiation emitted by charged particles produced in the interactions of neutrinos in the surrounding ice or the nearby bedrock. 
But IceCube is also sensitive to cosmic-ray properties by investigating high-energy muons originating from cosmic-ray air showers. 
Since the direction of atmospheric muons with multi-TeV energies is on average within $0.1^{\circ}$ of the direction of the original cosmic ray \cite{PhysRevD.89.102004}, directional analyses are possible as well.
These muons occur at a rate that is about a factor $10^{6}$ larger compared to atmospheric neutrinos and a factor of $10^{10}$ more than the cosmic neutrino flux. 
Neutrino signatures only become visible via the application of elaborate analysis methods. 
The large wealth of cosmic-ray data have been used in the past to study cosmic-ray anisotropies \cite{icecube_anisotropy,icecube_hawc2019} and the diffuse cosmic ray flux and composition \cite{aartsen_icetop2013,AndeenPlum:2019}. A first detection of the cosmic-ray Sun shadow with IceCube has been published in \cite{bos2019}. It was shown that even above an energy of 40 TeV, the cosmic-ray shadow of the Sun is not constant in time. It is the aim of this update of the analysis presented in \cite{bos2019} to investigate if the temporal variations are correlated with the magnetic field cycle.

In this analysis, we use IceCube data from the seasons May 2010 to May 2017 - the first season is labeled \textit{IC79} in the following, as only 79 strings were deployed until April 2011. The following seasons are labeled \textit{IC86-I to IC86-VI}, referring to the full IceCube configuration with 86 strings.

\section{7-year analysis of the cosmic-ray shadow of the Sun with IceCube}
\subsection{Sun- and Moon filters and quality cuts}
At trigger level, down-going muons, with an energy of roughly more than 400 GeV dominate the total trigger rate of 2,100 s$^{-1}$. Due to limited data transfer bandwidth of 100 GB per day from the South Pole to the Northern Hemisphere, online filters are used to reduce the amount of data. This analysis uses filtered data streams specifically designed to study the Moon and Sun shadows.
These filters are angular windows ($\pm 10^{\circ}$ in zenith, $\pm 180^{\circ}$ in azimuth) around the expected position of Moon and Sun. They are enabled when at least eight DOMs in three different strings detect photons. Furthermore, the Moon/Sun has to be above the horizon at the Pole for the IceCube seasons IC86-II and higher and at least $15^{\circ}$ above the horizon for the samples IC79 and IC86-I to satisfy the online filters.

The Sun reaches a maximum elevation of $\sim 23.5^{\circ}$ each year, while the Moon's maximum elevation varies between $\sim 24^{\circ}$ and $\sim 18^{\circ}$ at the geographic South Pole for the fraction of the lunar nodal precession covered by these observations. The Moon filter is enabled for several consecutive days each month, whereas the Sun filter collects data for approximately 90 days from November through February each season. At the South Pole, a fast likelihood-based track reconstruction is performed for each muon event and its direction is compared to the expected position of the celestial body in angular coordinates. The data are transferred to the Northern Hemisphere by satellite if the reconstructed event satisfies the Moon or Sun filter.
With this filtering, the data enter the standard level-2 data reduction within IceCube. 

Afterwards, further event cleaning is performed by applying quality cuts to remove events that are mis-reconstructed or have poor angular resolution. 
This analysis makes use of two cut variables: the angular uncertainty $\sigma$ of the reconstructed event and a track reconstruction quality estimator, the \textit{reduced log-likelihood} (rlogl). 
Both cut variables are typically used in point source and other IceCube analyses \cite{Abbasi_2011}. Assuming Poisson statistics, the significance $\tilde{S}$ of the shadowing effect is proportional to the square root of the fraction $\eta$ of events passing the cuts and the resulting median angular resolution $\sigma_{\rm med}$ after cuts, $\tilde{S}\propto \sqrt{\eta}/\sigma_{\rm med}$. For both, IC79 and IC86 data, these quality cuts are optimized in order to maximize the significance $\tilde{S}$. The resulting optimal cut values are $\sigma <0.71^{\circ}$ and $\mathrm{rlogl}<8.1$.

\subsection{Preparing the final sample}
The parameters of the final sample are given as follows: in an angular window with a size of $[54\times 6]$ degrees in right ascension and declination around the position of the Moon and the Sun, $(3.8-7.9)\cdot 10^{6}$ events for the Moon and $(9.0-13.3)\cdot 10^{6}$ events for the Sun pass the quality cuts of the analysis in each year. The number of events used in the Moon shadow analysis varies because of the change of the maximum elevation of the Moon each year. At higher elevations more muons reach the IceCube detector. For the Sun, there is only a difference between the IC79 and IC86 detector configurations.

In order to determine the necessary parameters used for the simulations, the average elevation of all events in the final sample of each year is calculated together with the weighted average of the apparent radius $\left<R_{\rm app}^{\rm Moon/Sun}\right>$. In doing so, the number of events for each MJD, $N_{\rm MJD}$ is determined together with the apparent radius on the sky for each individual MJD of the sample, $R_{\rm app}^{\rm MJD}$. The latter is calculated by obtaining the Earth-Moon distance for that specific MJD and applying trigonometry. Using these pieces of information, the weighted apparent radius becomes
\begin{equation}
\left<R_{\rm app}\right>=\frac{\sum_{\rm MJD}N_{\rm MJD}\,R_{\rm app}^{\rm MJD}}{\sum_{\rm MJD}N_{\rm MJD}}\,.
\end{equation}
With this procedure, the apparent radius of the Moon is shown to vary between $0.251^{\circ}$ and $0.274^{\circ}$ on the sky, while the Sun's radius only varies on a sub-percent level and is therefore assumed to be constant at $0.271^{\circ}$.

In order to prepare the data for analysis, relative coordinates are introduced for each event, given as the direction in right ascension and declination of the muon relative to the Moon or Sun, $\Delta\alpha=\alpha_{\mu}-\alpha_{\rm Moon/Sun}$ and $\Delta\delta=\delta_{\mu}-\delta_{\rm Moon/Sun}$, respectively. 
\begin{figure}
    \centering
    \includegraphics[width=0.9\linewidth]{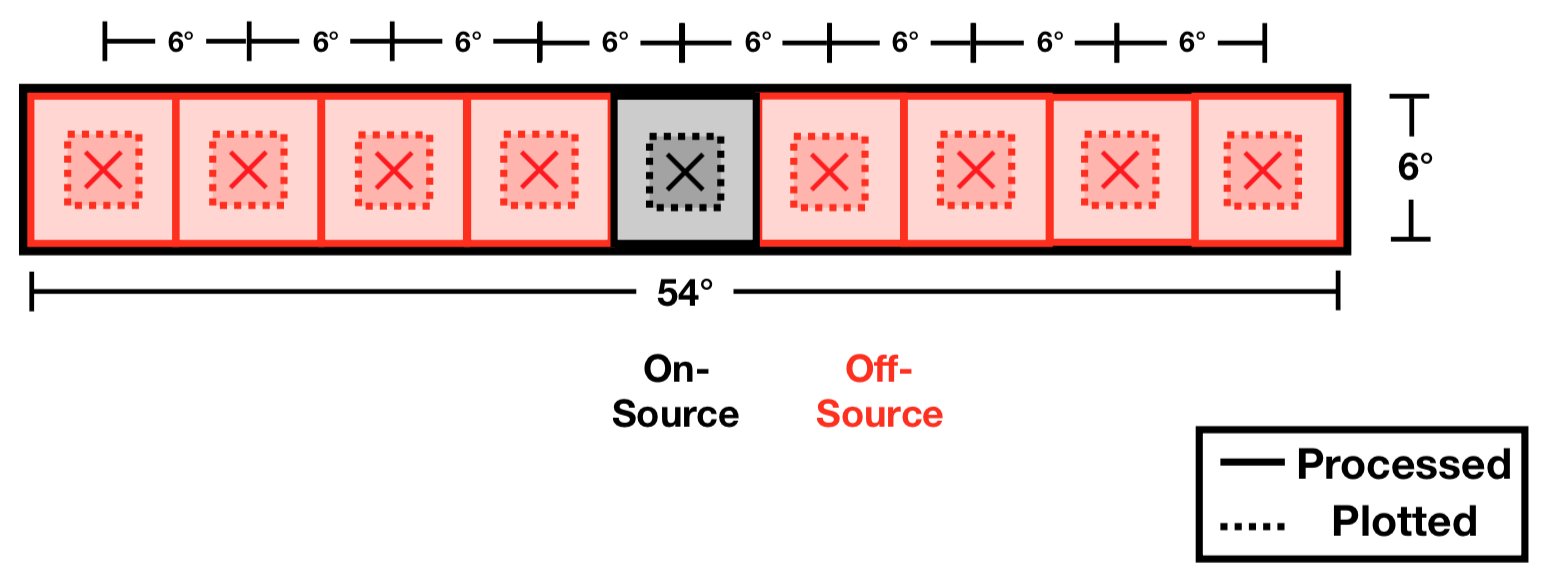}
    \caption{Layout of the analysis with 8 off-source bins for background determination (red) and one on-source bin (black/grey). The dashed boxes represent the region that is shown in the plots in Section \ref{results:sec}.}
    \label{sketch:fig}
\end{figure}
For the analysis, one on- and eight off-source windows with a size of $6^{\circ}\times 6^{\circ}$ are defined. A schematic view of the geometry of these windows is given in Fig.\ \ref{sketch:fig}. The figure shows the off-source regions in red and the on-source region in black. The dashed squares represent the region that is plotted when the results are shown in the next section.

First, the relative deficit is shown on a 2D grid in order to visualize the shadow and find its center of gravity. 
Then, the relative deficit in a circle with radius $1^{\circ}$ around the center of gravity is calculated as a measure of the shadowing effect.
For the 2D analysis, two two-dimensional maps are created; one map for the on-source and one map for the off-source regions. The axes are defined as $\cos\delta_{\mu}\cdot \Delta\alpha$ (x-axis) and $\Delta\delta$ (y-axis)). The $\cos\delta_\mu$ factor on the x-axis accounting for the distortion due to the spherical coordinate system assures equal solid angle. Each map is $3^{\circ}\times 3^{\circ}$ large and consists of $30\times 30$ bins, each $0.1^{\circ}\times 0.1^{\circ}$ large. 
Then, the relative deficit is calculated in each $0.1^{\circ}\times 0.1^{\circ}$ bin as follows:
\begin{equation}
    \frac{\Delta N}{\left<N\right>}=\frac{N_{\rm on}-\left<N_{\rm off}\right>}{\left<N_{\rm off}\right>}\,.
\end{equation}
Here, $N_{\rm on}$ is the number of events in the on-source region, and $\left<N_{\rm off}\right>$ is the average value in the eight off-source regions.
Due to the comparably low statistics per bin, it is necessary to apply a smoothing. A boxcar/top-hat smoothing with respect to angular distance is therefore performed. The relative deficit in each bin $(i,j)$ is replaced by the average of all bins for which the bin centers are within a certain angular distance around the center of bin $(i,j)$. Here, the smoothing radius is chosen to be $0.7^{\circ}$. 
This value approximately corresponds to the median angular resolution of the event sample.
After the smoothing procedure described above, the center of gravity of the shadow can be determined by averaging over the positions of the bins' centers for which the relative deficit after smoothing is greater than 3\%. Since the uncertainty in each bin amounts to about 0.5\%, this means only bins with a statistically significant deficit are taken into account. For the numerical analysis, the relative coordinates of each event are corrected in a way such that they are relative to the center of gravity and not relative to the expected position of Moon/Sun. The goal behind this \textit{center-of-gravity correction} is to separate out a possible systematic shift of the shadow due to the geomagnetic or solar magnetic fields, since in this work we are only interested in the total shadowing effect.

The numerical analysis of the shadow is then performed in two steps:
(1) The center-of-gravity correction is applied to each event; (2) The relative deficit is calculated in a circle around the center of gravity of the shadow. As the relevant radius, we choose $1^{\circ}$. When choosing the appropriate radius, there is a trade-off between the statistical error and capturing more shadowed events on the one hand and decreasing the sensitivity due to more and more background containment on the other hand. At $1^{\circ}$, the cumulative point spread function reaches approximately $68\%$ containment and the sensitivity of the relative deficit is still good enough.

The relative deficit within the $1^{\circ}$ search radius in $\%$ can be expressed as
\begin{equation}
    \mathrm{RD} (1^{\circ})=100\%\cdot \frac{N_{\rm on}(1^{\circ})-\left<N_{\rm off}(1^{\circ})\right>}{\left<N_{\rm off}(1^{\circ})\right>}\,.
    \label{deficit:equ}
\end{equation}

\section{Simulation of expected signal}
As a starting point, CORSIKA-based simulations \cite{corsika} covering primary energies from 600 GeV to 100 EeV and the five element groups H, He, CNO, MgAlSi and Fe, and based on the IC86 detector configuration, are processed the same way as the experimental data.
Weights are calculated according to the Hillas Gaisser model with a mixed extragalactic population \cite{2012APh....35..801G}.
The number of Monte Carlo (MC) generated events in these simulations ($\sim 5 \times 10^5$ events for a typical $\pm 3^{\circ}$ declination band) is statistically insufficient in order to perform the same analysis that is applied to the data. 
Generating a sufficient amount of simulated events is not feasible, as Monte-Carlo simulations are expensive, with respect to computing-time.
In order to increase the number of events to a reasonable amount, a resampling is performed.
For the resampling, each event is duplicated $20-100$ times, depending on the number of events in the specific declination band, and is assigned a random value between $-3$ and $+3$ degrees for $\Delta\alpha\cos\delta_\mu$. 
The relative right ascenssion of the corresponding primary cosmic rays are then shifted by the same amount as the random values are shifted compared to the originally reconstructed values.
This way, the difference in the relative right ascension between reconstructed direction and primary cosmic-ray direction is preserved.
The value for $\Delta \delta$ is kept in order to preserve the declination dependency of the sample.

For the simulation of the expected flux arriving at the IceCube detector, we use the concept of back-tracking particles 
by numerically integrating the equation of motion of nuclei in the solar magnetic field as described in \cite{sunshadow2019}. The back-tracking ensures that only those particles are considered that propagate in the direction of Earth after they have traversed the magnetic field (those particles, in turn, are relevant for the cosmic-ray Sun shadow observed at Earth). 

For the solar coronal magnetic field, two different potential field models are considered: the original version, called \textit{Potential Field Source Surface (PFSS)} model, introduces a magnetic potential by assuming a current-free surface, $\nabla \times B =\vec{0}$ \cite{Altschuler-Newkirk-1969,Schatten-etal-1969}. 
The \textit{Current Sheet Source Surface} (CSSS) model \cite{Zhao-Hoeksema-1995,Arge-Pizzo-2000} also takes into account current sheets which are neglected in the PFSS model.
Details of the theoretical modeling can be found in \cite{sunshadow2019}. 
By modeling this time-dependent magnetic field with its 11-year cycle structure, the theoretically predicted cosmic-ray shadow of the Sun is expected to change with time \cite{sunshadow2019}.

Differences as compared to \cite{sunshadow2019} are as follows: a) beyond the domain of the potential field models (here $2.5\,R_{\odot}$), the Parker spiral is considered using a simple analytical model of the solar wind velocity profile; b) the back-tracking is done for the particles coming out of the detector-specific CORSIKA-based simulations described above instead of particles with fixed energies and mass numbers; c) the back-tracking starts at the IceCube detector.

Then, the weights of the simulated events are modified with the goal to mimic the shadowing effects of the Moon and the Sun.
This is achieved by introducing a passing probability $p$ for each primary cosmic-ray particle that is assigned a value between 0 and 1 which is multiplied with the original weight $w$, yielding the modified weight $w' = w \cdot p$. 
For the Moon, $p$ either becomes 0 (primary cosmic-ray direction on Moon disk) or 1 (primary cosmic-ray direction not on Moon disk), while accounting for the changing size of the lunar disk by using $\left<R_{\rm app}\right>$.
For the Sun, $p$ is determined by back-tracking each initial particle in the magnetic field of the 4 to 5 solar rotations (Carrington rotations) that cover the observation time for the Sun.
A step-by-step rotation of the solar magnetic field for each Carrington rotation ensures that the simulation accounts for solar rotation during the observation period.
The passing probability $p$ of each event can then be calculated by dividing the number of cases in which the primary particle does not hit the Sun by the number of all cases.

\section{Results\label{results:sec}}
\begin{figure}
    \centering
    \includegraphics[width=0.55\linewidth]{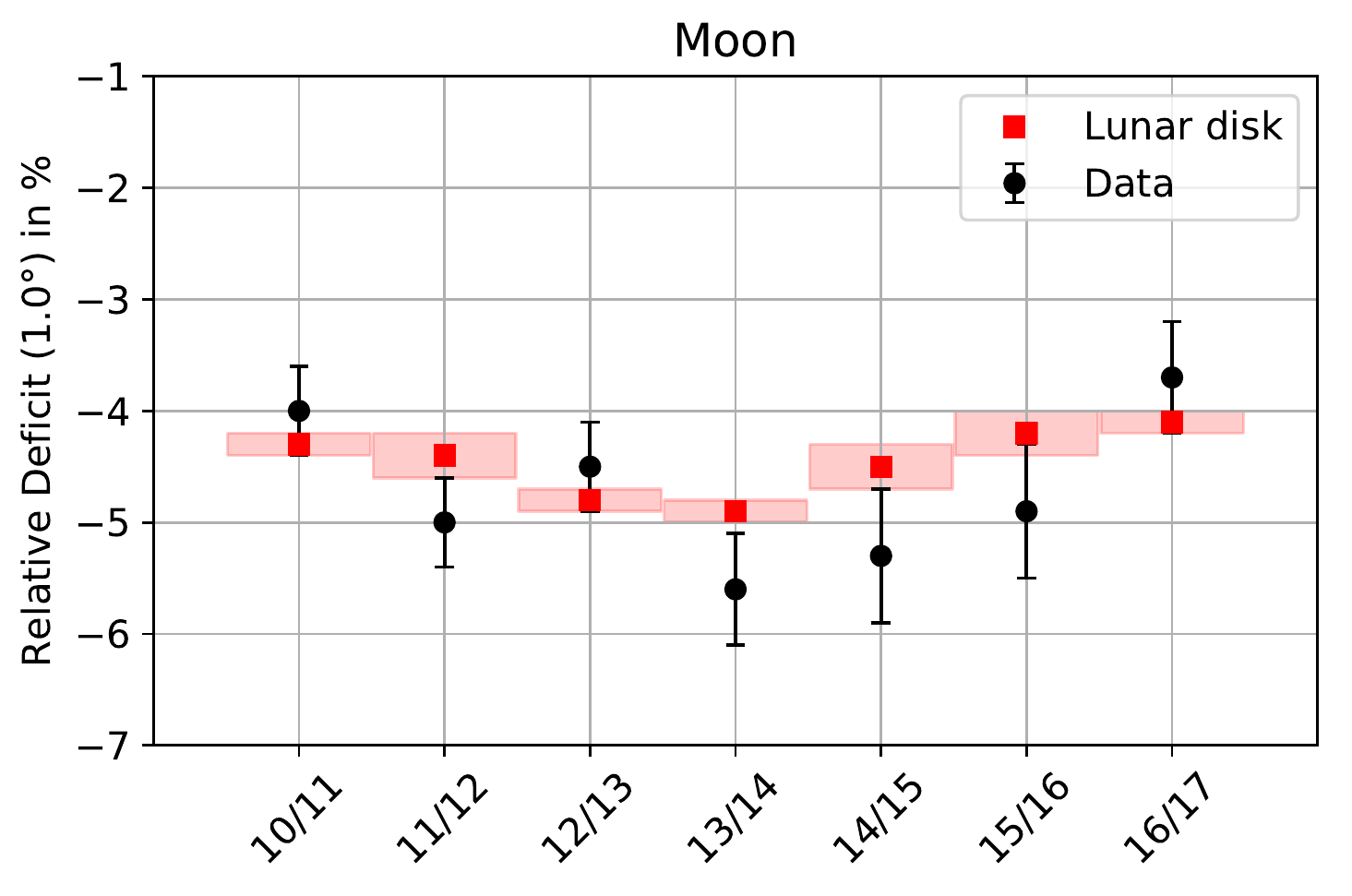}
    \caption{Relative deficit of the  Moon as a function of time.}
    \label{moon_depth:fig}
\end{figure}
\begin{figure}
    \centering
    \includegraphics[width=0.55\linewidth]{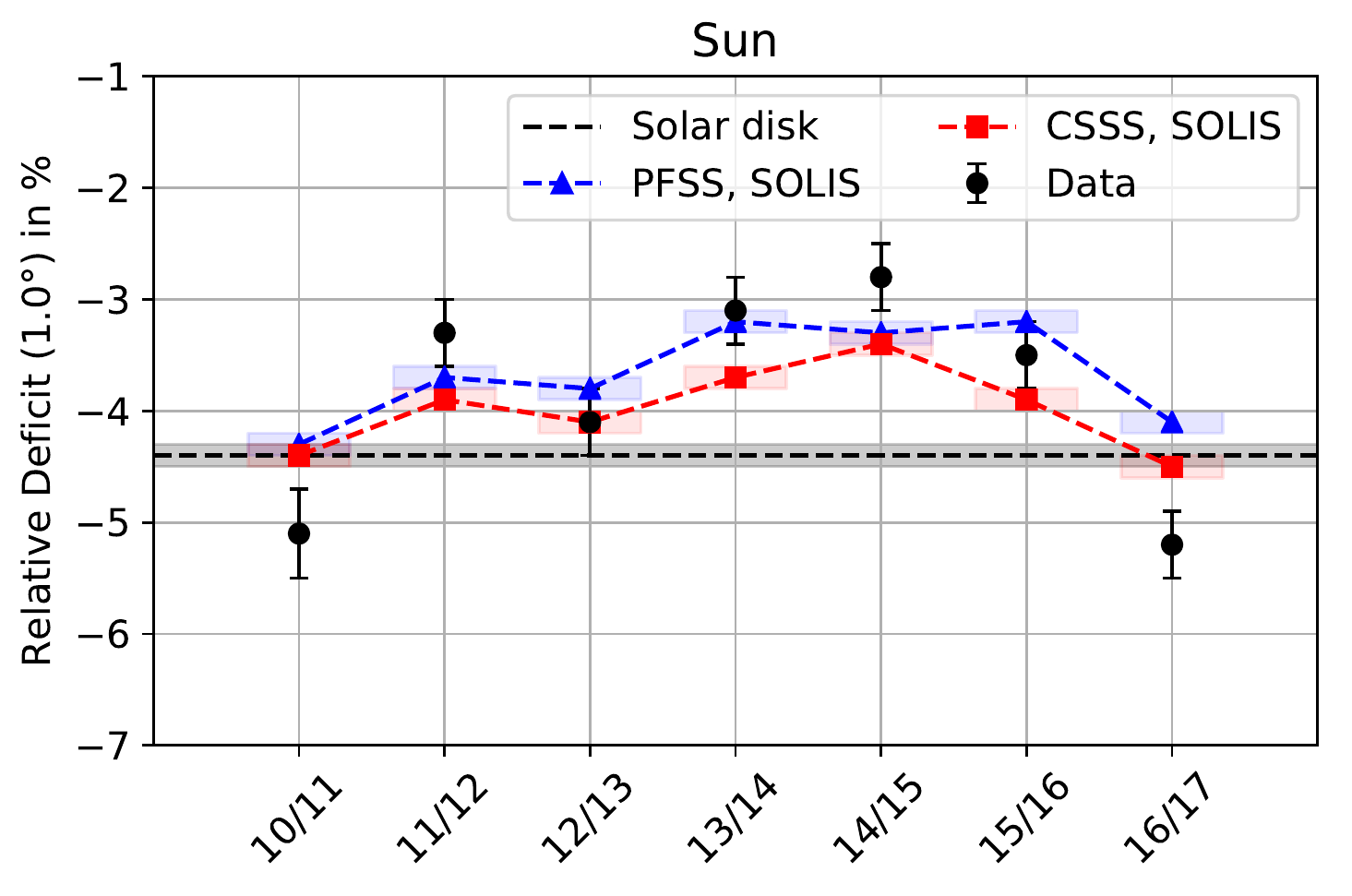}
    \caption{Relative deficit of the Sun as a function of time. Magnetogram input data for both models were acquired by SOLIS instruments operated by NISP/NSO/AURA/NSF.}
    \label{sun_depth:fig}
\end{figure}
The results for the relative deficit according to Eq.\ \ref{deficit:equ} for the Moon as a function of time are shown in Fig.\ \ref{moon_depth:fig}. Red squares represent the results for the simulation of the lunar disk, black circles show the data. Within $1\sigma$, no deviations from the expectation of a lunar disk can be observed (p-value of 0.32). 
Thus, the Moon behaves as expected according to the assumption that cosmic-rays are blocked. 
For the Sun, the results are shown in Fig.\ \ref{sun_depth:fig}. 
The expectation of a solar disk is represented by the black, dashed line, data are shown as black circles. 
In this case, the deviation from a simple model of cosmic-ray interactions in the Sun can be rejected with a p-value of $3.9\cdot10^{-13}$, i.e.\ a significance of $7.3\sigma$. Simulation results assuming cosmic-ray propagation in the PFSS and CSSS magnetic field models are shown as blue upward triangles and red squares, respectively. 
These model predictions are not fully compatible with the data ($3.0\sigma$ deviation for the PFSS model and $2.8\sigma$ for the CSSS model), but these deviations are small compared to the one of the solar disk. 
\begin{figure}
    \centering
    \includegraphics[width=0.55\linewidth]{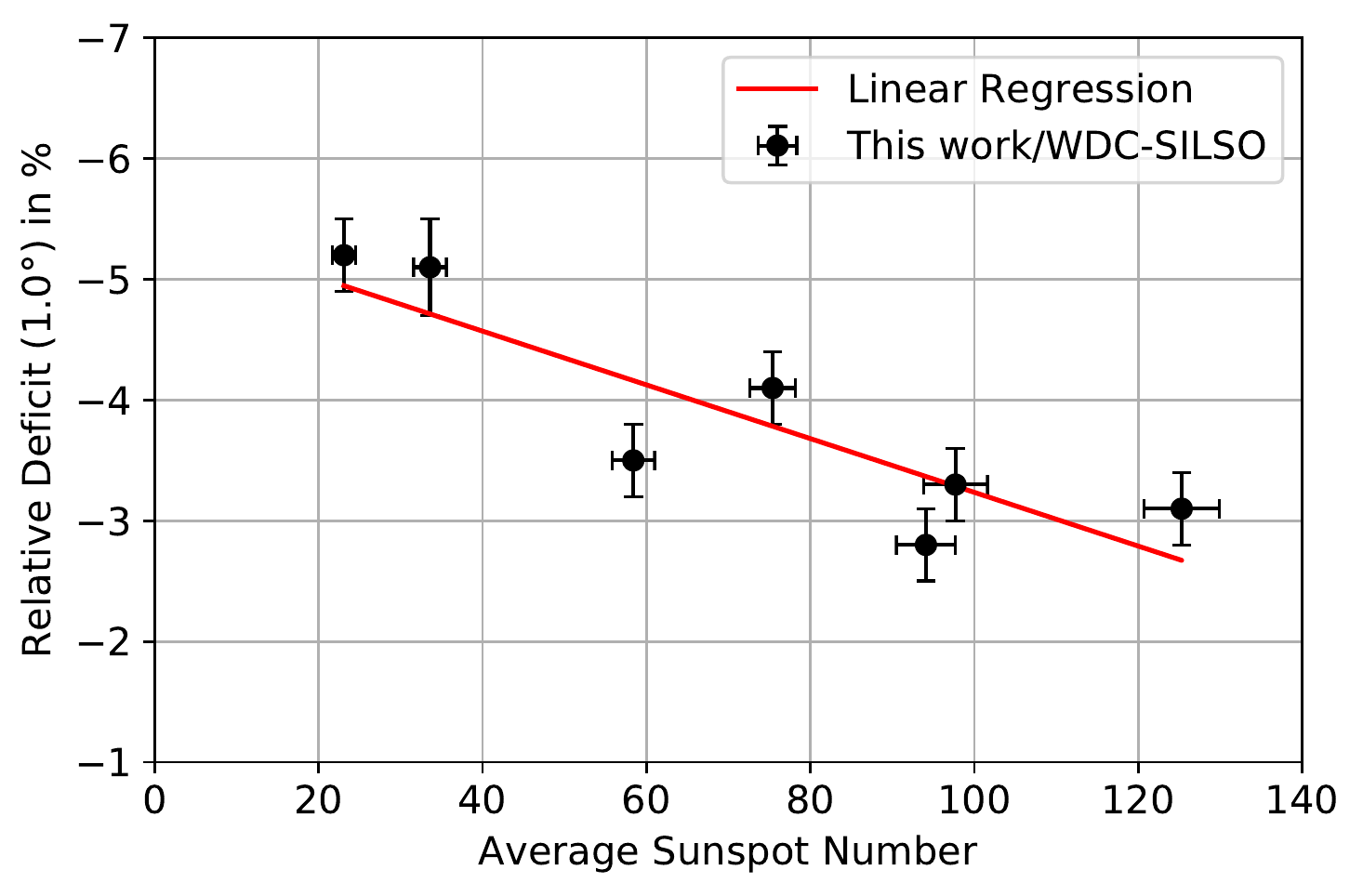}
    \caption{Average Sunspot number versus relative deficit. Sunspot data from the World Data Center SILSO, Royal Observatory of Belgium, Brussels \cite{sidc}.}
    \label{sun_spots:fig}
\end{figure}
Figure \ref{sun_spots:fig} shows the relative deficit as a function of the average sunspot number. 
%A spearman test has been performed, showing a clear correlation with a coefficient of $\rho=0.86$. Even with Kendall's test, a correlation can be shown to be present on a statistical significant level ($\kappa =0.71$). 
The data favor a (decreasing) linear fit over a constant fit with a significance of $6.4\,\sigma$. These numbers, independent on the exact solar field model prove a variation with solar activity.

\section{Conclusions and outlook}
We show that the cosmic-ray shadow of the sun varies in agreement with the 11-year solar cycle as indicated by the correlation to the sunspot number.
Using IceCube data with a median energy of $50$--$60$~TeV, the PFSS and CSSS models clearly show a better agreement than modelling the Sun as a disk. 
However, there are still tensions at the $3\,\sigma$ level with both of these models. 
Reasons could be that the simulations performed only take into account the local magnetic field model around the Sun in the Parker spiral approximation based on a simplified model of the solar wind velocity and that the impact of CMEs is not taken into account. The magnetogram measurements used as an input for the solar magnetic field modeling have a systematic error in the absolute scale of the magnetic field strength as they differ amongst each other by a factor of $\sim 2$ \cite{Riley2014}. The modeling via potentials is further only an approximation and might not describe reality precisely enough.
Using future measurements with IceCube and other cosmic-ray observatories will help answering the above questions. In particular, 
\cite{sunshadow2019} predict that the energy-dependent investigation of the Sun shadow opens the opportunity to test the solar magnetic field in detail. 

%\section{References}\label{sec:refs}

%This is a paper from a previous ICRC \cite{Zoll:2015wcu}. This is a second paper from a previous ICRC \cite{Peiffer:2017vsm}. This is a paper from the current ICRC \cite{Hussain:2019icrc_gw}.
%Here is an IceCube journal paper \cite{Aartsen:2016nxy} and an external journal paper \cite{Waxman:1998yy}.

% Set up the bibliography using BibTeX.
% Get references from inspirehep.net or NASA/ADS and put them in references.bib.
\bibliographystyle{ICRC}
\bibliography{icrc_sunshadow}

% Or, set up the bibliography manually, if you prefer to do things this way.
%
% \begin{thebibliography}{99}
%   \bibitem{Zoll:2015wcu}{{\bf IceCube} Collaboration, \pos{PoS(ICRC2015)1099} (2016).}
%   \bibitem{Peiffer:2017vsm}{{\bf IceCube-Gen2} Collaboration, \pos{PoS(ICRC2017)1052} (2018).}
%   \bibitem{Hussain:2019icrc_gw}{{\bf IceCube} Collaboration, \pos{PoS(ICRC2019)xyz} (these proceedings).}
%   \bibitem{Aartsen:2016nxy}{{\bf IceCube} Collaboration, M.~G.~Aartsen {et al.}, \emph{JINST} {\bf 12} (2017) P03012%
%   % optionally add arXiv ID here [{\tt astro-ph/1612.05093}]
%   .}
%   \bibitem{Waxman:1998yy}{E. Waxman and J. N. Bahcall, \emph{Phys. Rev.} {\bf D59} (1999) 023002.}
% \end{thebibliography}

\end{document}